\documentclass[a4paper]{jpconf}
\usepackage{graphicx}
\begin{document}
\title{Subbarrier fusion of carbon isotopes: from resonance structure to
fusion oscillations}

\author{K. Hagino$^{1,2}$ and N. Rowley$^3$}

\address{$^1$ Department of Physics, Tohoku University, Sendai 980-8578, Japan}
\address{$^2$ Research Center for Electron Photon Science, Tohoku University,
  1-2-1 Mikamine, Sendai 982-0826, Japan} 
\address{$^3$ 
  Institut de Physique Nucl\'{e}aire, UMR 8608, CNRS-IN2P3 et Universit\'{e}
de Paris Sud, 91406 Orsay Cedex, France}

\begin{abstract}
At energies below the Coulomb barrier, the fusion excitation 
function for the $^{12}$C+$^{12}$C system shows prominent fine structures, whereas 
that for the $^{12}$C+$^{13}$C system behaves more
smoothly as a function of energy. 
We demonstrate that these different
behaviors can be simultaneously reproduced 
using an optical potential in which the strength of the imaginary part 
is proportional to the 
level density of each compound nucleus. 
We also discuss the oscillatory behavior of fusion excitation function 
for these systems observed at energies above the Coulomb barrier from a
view point of quantum mechanical systems with identical particles. 
\end{abstract}

\section{Introduction}

The $^{12}$C+$^{12}$C fusion reaction plays an important role in several 
astrophysical phenomena, such as the carbon burning in stellar evolution, 
type Ia supernovae, and the X-ray superburst of an accreting neutron star. 
Even though there is a long history of research on this reaction, 
the reaction still attracts lots of attention and new experimental and 
theoretical works have continuously been carried out
~\cite{Spillane07,Notani12,Jiang13,Esbensen11,Assuncao13}. 
A characteristic feature of 
the $^{12}$C+$^{12}$C fusion reaction is that 
the cross sections have many fine structures, at energies both below and above 
the Coulomb barrier, while the cross sections for the neighboring 
systems, $^{12}$C+$^{13}$C and $^{13}$C+$^{13}$C, are much less structured. 
In this contribution, we simultaneously analyze fusion reactions for the 
$^{12}$C+$^{12}$C and $^{12}$C+$^{13}$C systems and discuss 
possible
origins for the different behavior of fusion excitation functions 
for these systems.
Notice that most of the previous studies, except for Ref. \cite{Esbensen11}, 
have concentrated only on the $^{12}$C+$^{12}$C system. 
In contrast, we shall analyze both the 
$^{12}$C+$^{12}$C and $^{12}$C+$^{13}$C systems 
and clarify the dynamics of 
subbarrier fusion of two carbon isotopes. 

\section{Subbarrier molecular resonances}

We first discuss the resonance behavior of fusion cross sections 
for the $^{12}$C+$^{12}$C system at energies below the Coulomb barrier. 
Figure 1(a) shows the experimental data for the modified
astrophysical $S$-factor, defined as $S^*(E)=\sigma(E)E\,
\exp(87.21/\sqrt{E}+0.46E)$,
where $\sigma(E)$ is the fusion cross section and the energy $E$ is in units
of MeV, for the $^{12}$C+$^{12}$C and the $^{12}$C+$^{13}$C
systems~\cite{Notani12}. 
As has been well known~\cite{Bromley60}, the fusion cross sections for
the $^{12}$C+$^{12}$C system exhibit a few resonance peaks.
Those relatively narrow
resonances are known as molecular resonances, for which the excitation
of $^{12}$C to the first 2$^+$ state plays an
important role~\cite{Imanishi68,Kondo78,Scheid72}.

\begin{figure}[htb]
\begin{center}
  \includegraphics[clip,scale=0.5]{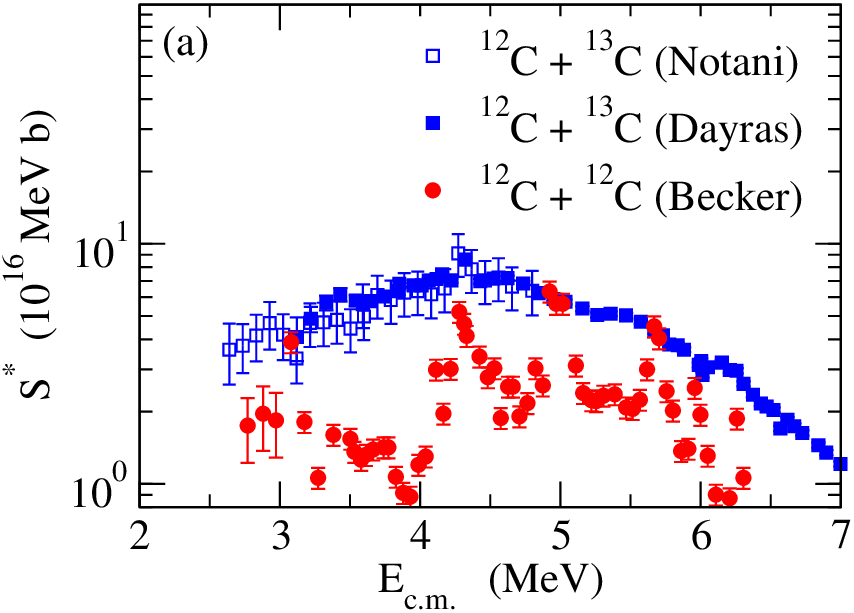}
  \includegraphics[clip,scale=0.5]{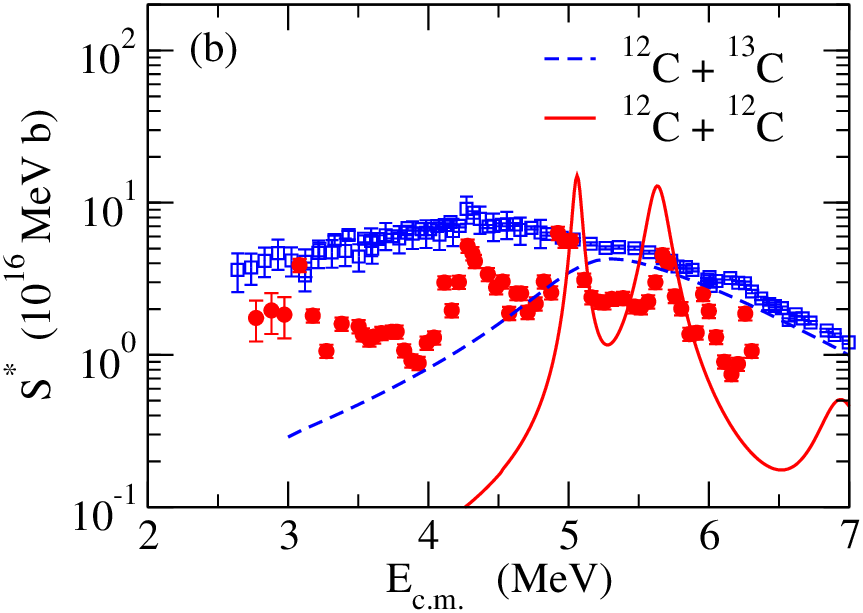}
\end{center}
\caption{(The left panel)
  The experimental data for the modified astrophysical $S$-factor
  for the $^{12}$C+$^{12}$C system (the filled circles) and for
  the $^{12}$C+$^{13}$C system (the filled and open squares).
The experimental data are taken from Refs.~\cite{Notani12,Becker81,Dayras76}.
(The right panel) Same as the left panel, but with
  theoretical curves for the $^{12}$C+$^{12}$C system (the solid line) and for
  the $^{12}$C+$^{13}$C system (the dashed line) obtained with the
  level density dependent optical potentials.
}
\end{figure}

As has been pointed out in Ref.~\cite{Notani12}, the fusion cross sections
for the $^{12}$C+$^{12}$C system
show inhibitions as compared to those for
the $^{12}$C+$^{13}$C system except at a few resonance energies, at which
the fusion cross sections for the two systems somehow match with each other.
Recently, Jiang {\it et al.} have argued that the fusion inhibition in
the $^{12}$C+$^{12}$C system 
is attributed to
i) the smaller fusion $Q$-value in the entrance channel
compared to that for the $^{12}$C+$^{13}$C and $^{13}$C+$^{13}$C systems, 
ii) a smaller level density of the compound nucleus $^{24}$Mg
than $^{25}$Mg and $^{26}$Mg at a given excitation energy,
and iii) the fact that
only states with positive parity and even spin of the compound
nucleus are populated in fusion because the entrance channel consists
of identical spin-zero bosons~\cite{Jiang13}.
Jiang {\it et al.} have succeeded to explain the {\it average} behavior
of fusion cross sections for the $^{12}$C+$^{12}$C system based on this idea.

Our first
aim in this contribution is to implement the idea of Jiang {\it et al.}
into coupled-channels calculations and discuss the difference
between the $^{12}$C+$^{12}$C and $^{12}$C+$^{13}$C systems.
To this end, we carry out coupled-channels calculations with
an optical potential, in which the strength of the imaginary part is
proportional to the level density of the compound nucleus.
That is, the imaginary part of the optical potential is assumed to be
\begin{equation}
W(r)=-w_0\rho_J(E^*)f(r),
\label{Impart}
\end{equation}
where $w_0$ is an overall strength, $\rho_J(E^*)$ is the level
density of the compound nucleus at the excitation energy of $E^*$
with the angular momentum $J$, and $f(r)$ determines the radial dependence
of the imaginary potential, which we assume to be a Woods-Saxon form.
The level-density dependent imaginary potential has been employed 
in Refs.~\cite{Scheid72,Helling71}, which has further been 
investigated in Refs.~\cite{Quesada83,Andres85}.
Eq. (\ref{Impart}) may be justified in terms of the Fermi's golden rule for
a transition from the entrance channel to compound nucleus
states~\cite{Scheid72,Helling71,Quesada83,Andres85}.
In this approach, 
the energy, the angular
momentum, and the system dependences of the imaginary potential
are taken into account through the level density of the compound nucleus. 

In order to perform the coupled-channels calculations for the
$^{12}$C+$^{12}$C and $^{12}$C+$^{13}$C systems, we closely follow the calculations
presented in Ref.~\cite{Kondo78}. That is, we use the same geometry for the
optical potential as in Ref.~\cite{Kondo78}, and we include
the excitations of both the projectile and the target nuclei up to the
first member of the ground state rotational band (including also
the mutual
excitation channels). 
We take the deformation parameters from Ref.~\cite{Esbensen11}, while
we take the empirical
level density parameters from Ref.~\cite{Jiang13}.
The overall strength $w_0$ in Eq. (\ref{Impart}) is determined so as to
reproduce the results of Ref.~\cite{Kondo78} if all the parameters are set
identical to those in Ref.~\cite{Kondo78}.
We use the same value of $w_0$ both for the 
$^{12}$C+$^{12}$C and $^{12}$C+$^{13}$C systems.
Most of the channels are closed, and the coupled-channels equations
are solved with the variational method~\cite{Kondo78,Matsuse78,Kamimura77} 
in order
to avoid a numerical instability. 

Figure 1(b) shows the results so obtained. The solid and the dashed lines
show the modified astrophysical $S$-factors for the $^{12}$C+$^{12}$C
and the $^{12}$C+$^{13}$C systems, respectively.
One can see that the system dependence of fusion cross sections is
qualitatively well reproduced. That is, the calculated $S$-factors for 
the $^{12}$C+$^{12}$C system show resonance peaks, while those for the
the $^{12}$C+$^{13}$C system behave rather smoothly. This indicates that
a promising origin for the resonance behavior in 
the $^{12}$C+$^{12}$C system is attributed to
the properties of the compound nucleus $^{24}$Mg,
as has been suggested by Jiang {\it et al.}~\cite{Jiang13}.

The calculations still underestimate the fusion cross sections
below 5 MeV for both the systems.
This would be due to those channels which are not included in the
present calculations. 
Possible candidates are the first 3$^-$ state and
the second 0$^+$ state (that is, the Hoyle state)
~\cite{Esbensen11,Assuncao13}. The $\alpha$-transfer channel may also play
an important role~\cite{Cujec89,Wada77,Wada80}. 
It would be an interesting future work to repeat the present calculation 
by including those channels.

\section{Fusion oscillations above the Coulomb barrier}

We next discuss fusion cross sections at energies 
above the Coulomb barrier. 
The experimental data for the $^{12}$C+$^{12}$C system 
are shown in Fig. 2 (a). 
One can see that the fusion cross sections significantly 
oscillate as a function of energy. 
The origin for the fusion oscillations may be different 
from the origin for the resonance behavior at subbarrier energies
shown in Fig. 1, however. 

\begin{figure}[tb]
\begin{center}
\includegraphics[clip,scale=0.5]{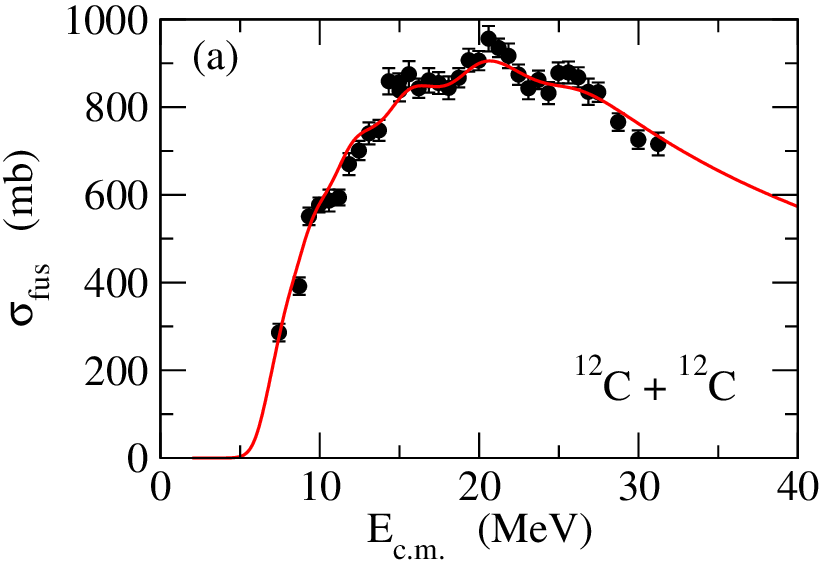}
\includegraphics[clip,scale=0.5]{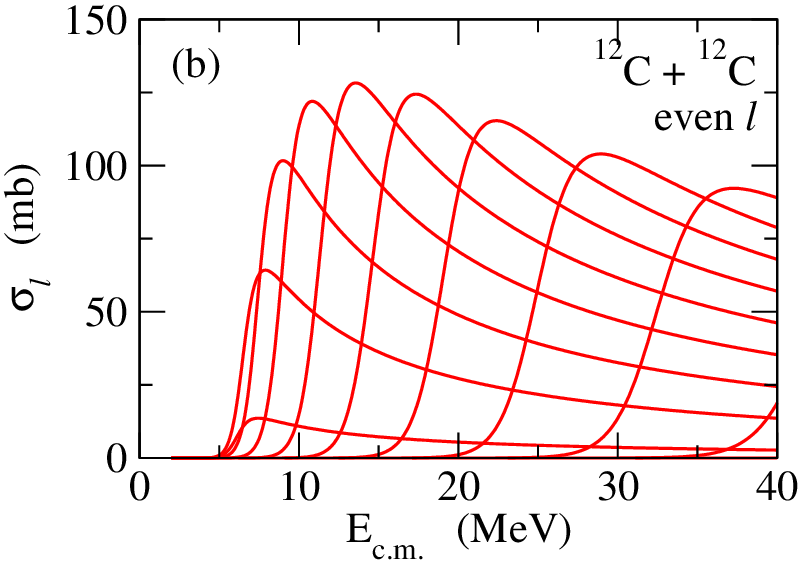}
\end{center}
\caption{(The left panel) Comparison of 
the experimental fusion cross sections  
to the result of optical model calculation 
for the $^{12}$C+$^{12}$C system. The experimental data 
are taken from Ref.~\cite{Kovar79}. (The right panel) The partial 
fusion cross sections corresponding to the theoretical 
result shown in Fig. 2(a). }
\end{figure}

At energies above the barrier, the level density 
is large enough so that resonances in the compound nucleus appreciably 
overlap with each other. This results in the picture of strong absorption, 
with which fusion cross sections are given by,
\begin{equation}
\sigma(E)=\frac{\pi}{k^2}\sum_l(2l+1)P_l(E)=\sum_l\sigma_l(E),
\end{equation}
where $k$ is the wave number related to the incident energy in the 
center of mass frame, $P_l$ is the penetrability of the Coulomb 
barrier with the partial wave $l$, and $\sigma_l(E)$ is a partial 
fusion cross section. 
For each partial wave, the penetrability $P_l(E)$ is close to 
unity and does not change significantly 
at energies well above the barrier. The partial cross section $\sigma_l(E)$ 
then decreases as $1/k^2$ as a function of energy. Since the partial waves 
are discrete, one would then obtain a structure 
in fusion cross sections, which are associated with the energy 
dependence of partial fusion cross sections. This is the origin for 
the fusion oscillations advocated in Refs.~\cite{Poffe83,Kabir88} 
(see also Ref. \cite{Esbensen12}). That is, the fusion oscillations are 
due to the addition of successive individual partial waves as the energy 
increases. 

For the $^{12}$C+$^{12}$C system, since the 
wave function of the whole system has to be symmetric with respect to 
the interchange of two identical spin-zero
bosons, only even partial waves contribute 
to fusion cross sections. That is, 
\begin{equation}
\sigma(E)=\frac{\pi}{k^2}\sum_l(1+(-1)^l)(2l+1)P_l(E)=\sum_l
(1+(-1)^1)\sigma_l(E). 
\end{equation}
This leads to an enhancement of oscillations, since 
the energy spacing between successive 
contributing partial waves then increases. 
 
The solid line in Fig. 2(a) shows a potential model fit to the experimental 
fusion cross sections. Since the effect of channel couplings are small for 
this system at energies above the Coulomb barrier, 
we perform 
single-channel calculations for simplicity. 
In order to account for the observed
decrease of fusion cross sections at energies 
higher than 25 MeV, we have reduced
the penetrability for $l=14$ by a factor of 2 
and set the penetrabilities for all the higher partial waves to be zero. 
Such an assumption may be justified from a consideration based on 
the excitation energy of the compound nucleus relative to the yrast 
energy~\cite{HR14,HR14-2}. This calculation 
fairly well reproduces the experimental 
data, including the fusion oscillations. 
The structure of fusion cross sections is evident if 
the corresponding partial fusion cross sections are plotted individually,
as is done in Fig. 2(b). 

Within the parabolic approximation to the Coulomb barrier,
one can derive a compact formula for 
the oscillatory part of fusion cross sections
\cite{Poffe83,HR14,HR14-2,HT12},
\begin{equation}
\sigma_{\rm osc}(E)=2\pi R_E^2\frac{\hbar\Omega_E}{E}\exp\left(-
\frac{\pi\mu R_E^2\hbar\Omega_E}{(2l_g+1)\hbar^2}\right)\sin(\pi l_g),
\label{osc}
\end{equation}
which is added to the smooth part of fusion cross sections given by the
well known Wong formula~\cite{Wong73}. 
In this equation,
$\mu$ is the reduced
mass of the system, and 
$R_E$ and $\hbar\Omega_E$ are the barrier position and the
curvature of the Coulomb barrier, respectively,
evaluated at the grazing angular momentum,
$l_g$. For a spatially anti-symmetric configuration, 
the negative sign has to be multiplied to this formula 
\cite{HR14,HR14-2}. 
This formula indicates that the amplitude of fusion oscillations exponentially
decrease for heavy systems (with large values of $\mu$), and one has a better
chance to see the fusion
oscillations in light systems, such as $^{12}$C+$^{12}$C and
$^{16}$O+$^{16}$O.

\begin{figure}[tb]
\begin{center}
\includegraphics[clip,scale=0.5]{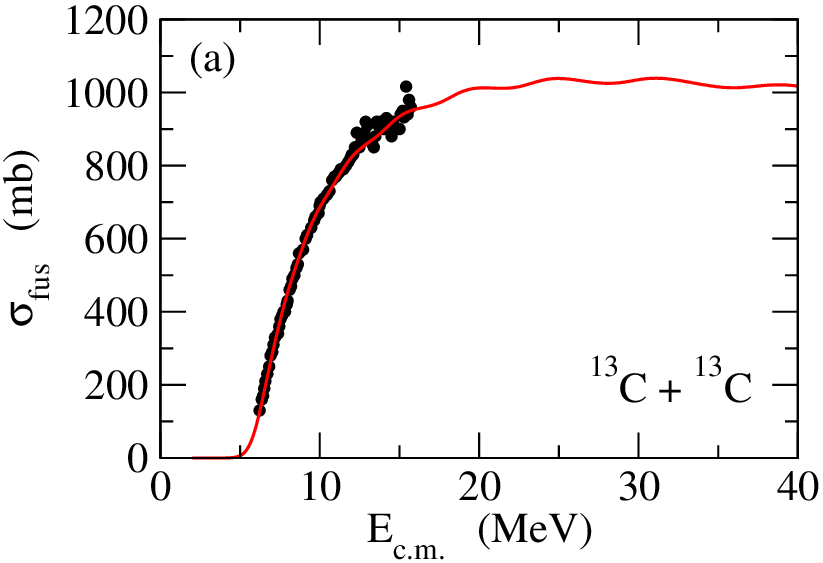}
\includegraphics[clip,scale=0.5]{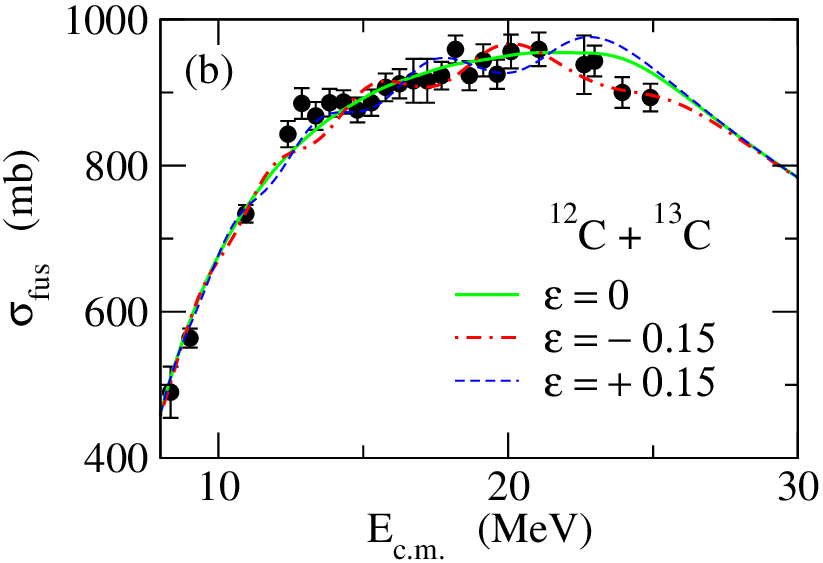}
\end{center}
\caption{(The left panel)
  Same as Fig. 2(a), but for the $^{13}$C+$^{13}$C system. The experimental
  data are taken from Ref.~\cite{Charvet82}.
  (The right panel) The experimental fusion cross sections for the
  $^{12}$C+$^{13}$C system taken from Ref.~\cite{Kovar79}.
  The three curves are obtained with a
  parity dependent potential with the depth parameter of
  $V=(1+(-1)^l\epsilon)V_0$, where $V_0$ is a negative value, with different
  values of $\epsilon$ as indicated in the figure. 
}
\end{figure}

The experimental fusion cross sections 
for the $^{13}$C+$^{13}$C are shown in Fig. 3(a).
One can see that the amplitude of fusion oscillation is much smaller than that
for the $^{12}$C+$^{12}$C system shown in Fig. 2(a).
The $^{13}$C+$^{13}$C system is a system of two identical spin-1/2 fermions,
and both the $S$=0 (which gives even $l$ with weight 1/4)
the $S$=1 (which gives odd $l$ with weight 3/4) configurations contribute.
Since the oscillatory cross section for odd partial waves has the opposite
sign to that for even partial waves, the oscillation is reduced by a factor
2 and has the opposite phase from the symmetric system. The solid line
in Fig. 3(a) is obtained in this way.

For the $^{12}$C+$^{13}$C system, since this is not a system with identical
particles, the oscillations from even $l$ and odd $l$ would cancel out
without any further effect. However, the presence of the elastic neutron
transfer channel will introduce a parity dependence into the problem.
This is most easily seen by considering the total elastic scattering, where
one must add an exchange term $f_{\rm trans}(\pi-\theta)$ for elastic
transfer to the 
amplitude $f_{\rm el}(\theta)$ for direct elastic scattering. 
This yields a total scattering amplitude,
$f_{\rm total}(\theta)=f_{\rm el}(\theta)+f_{\rm trans}(\pi-\theta)$.
Using $P_l(\cos(\pi-\theta))=(-1)^lP_l(\cos\theta)$,
one obtains $S_l^{\rm eff}=S_l^{\rm el}+(-1)^lS_l^{\rm trans} $,
that is, different effective $S$-matrix elements for the odd and even partial
waves. The same mechanism with $\alpha$ transfer has been discussed
in Ref.~\cite{Kabir88} for the fusion oscillations observed
in the $^{12}$C+$^{16}$O system.
As is well known, the parity dependence due to elastic transfer can be well
mocked up with a parity dependent potential
~\cite{Kabir88,vonOertzen75,Vitturi86,Christley95}. 
The three different lines in Fig. 3(b) are obtained with a parity dependent
potential, whose depth parameter is defined as $V=(1+(-1)^l\epsilon)V_0$ (with
a negative value of $V_0$). Treating $\epsilon$ as a free parameter,
we obtain a good fit to the experimental data with $\epsilon=-0.15$,
that is, a shallower potential (and thus a higher Coulomb barrier)
for even partial waves.
We mention that the sign of $\epsilon$
found for the $^{12}$C+$^{13}$C system 
is consistent with a simple rule
proposed by Baye based on the resonating group method (RGM) with a
two-center harmonic oscillator shell model~\cite{Baye77,Baye86}.

Notice that, around $E\sim$13 MeV, the calculation with 
$\epsilon=+0.15$ appears more consistent with the experimental data
compared with the result with $\epsilon=-0.15$. 
It would be interesting to remeasure fusion cross sections in this 
energy region with higher precision to clarify whether 
there indeed exists a possible shift in phase of the oscillations 
as a function of energy.

\section{Summary}

The fusion reaction of carbon isotopes is important from
the astrophysical point of view, and at the same time it makes also an
interesting quantum mechanical problem.
We have discussed this reaction from the subbarrier
to the above barrier regions by comparing the $^{12}$C+$^{12}$C system to
the neighboring $^{12}$C+$^{13}$C and $^{13}$C+$^{13}$C systems. 

At subbarrier energies, the fusion cross sections for the 
$^{12}$C+$^{12}$C system show many resonance peaks whereas those for
the other systems behave smoothly. We have demonstrated that this fact
can be naturally explained if one considers properties of the compound
nuclei, as had been conjectured by Jiang {\it et al.}.
To this end, we have carried out coupled-channels calculations by
including excitations to the first excited state in the
ground state rotational band. In these calculations, we have employed 
an optical potential whose strength
is directly proportional to the level density of each compound nucleus.

At energies above the Coulomb barrier, the fusion cross sections
for the $^{12}$C+$^{12}$C system oscillate as a function of energy.
We have demonstrated that this oscillation can be interpreted as
due to the addition of successive individual partial waves. 
The oscillation is stronger in the $^{12}$C+$^{12}$C system than in
the $^{13}$C+$^{13}$C system, because only even partial waves contribute
to fusion cross sections for the former system while both
even and odd partial waves contribute (with
a statistical weight of 1:3) for the latter system.
We have also shown that the fusion oscillation observed in 
the $^{12}$C+$^{13}$C system is due to elastic transfer of neutron, whose
effect is well mocked up in terms of a parity dependent potential.

It still remains as a challenging problem to quantitatively
explain observed fusion cross sections for the C+C systems
at deep subbarrier energies.
In order
to achieve this goal, 
one would have to take into account the effects of several
collective excitations as well as the alpha transfer channel.

\ack
We thank X.D. Tang and S. Shimoura for useful discussions.
This work was supported by JSPS KAKENHI Grant Number 25105503.

\end{document}